\documentclass[useAMS,usenatbib]{mn2e}
\usepackage{natbib}
\usepackage{graphicx}
\usepackage{savesym}
\usepackage{listings}
\usepackage[intlimits]{amsmath}
\savesymbol{iint}
\usepackage{txfonts}
\usepackage{bm}
\restoresymbol{TXF}{iint}
\usepackage{amssymb}
\usepackage{hyperref}
\usepackage{amsmath}
\usepackage{tabularx} 
\usepackage{color}
\usepackage{soul}
\usepackage{ulem}

\def\apjl{Astrophys. J. Lett.}
\def\apjs{Astrophys. J. Supp. Ser. }

\def\aap{Astron. Astrophys. }

\newcommand{\sub}[1]{_{\mbox{\tiny #1}}}
\newcommand{\vv}[1]{{\bf #1}}

\newcommand{\Pd}[1]{\partial_{#1}}

\newcommand{\beq}{\begin{equation}}
\newcommand{\eeq}{\end{equation}}
\newcommand{\fracp}[2]{\left(\frac{#1}{#2}\right)}

\newcommand{\spr}[2]{{\bf #1} \!\cdot\! {\bf #2}}

\title[Magnetic inhibition of recollimation instability]{Magnetic inhibition of the recollimation instability in relativistic jets}
\author[J. Matsumoto et al.]{Jin Matsumoto$^{1}$ \thanks{E-mail: jin.matsumoto@fukuoka-u.ac.jp, jin@kusastro.kyoto-u.ac.jp},
Serguei S. Komissarov$^{2}$ \thanks{Email: S.S.Komissarov@leeds.ac.uk}
and Konstantinos N. Gourgouliatos$^{3}$ \thanks{Email: kngourg@upatras.gr}\\
$^{1}$Research Institute of Stellar Explosive Phenomena, Fukuoka University, Fukuoka 814-0180, Japan\\
$^{2}$School of Mathematics, Faculty of Mathematics and Physical Sciences, University of Leeds, Leeds, LS2 9JT, UK\\
$^{3}$Department of Physics, University of Patras, Patras 26504, Greece}

\begin{document}

\maketitle
\date{\today}

\begin{abstract}
In this paper, we describe the results of three-dimensional relativistic magnetohydrodynamic simulations aimed at probing the role of regular magnetic field on the development of the instability that accompanies recollimation of relativistic jets. In particular, we studied the recollimation driven by the reconfinement of jets from active galactic nuclei (AGN) by the thermal pressure of galactic coronas.   We find that a relatively weak azimuthal magnetic field can completely suppress the recollimation instability in such jets, with the critical magnetisation parameter $\sigma_{cr} < 0.01$. We argue that the recollimation instability is a variant of the centrifugal instability (CFI) and show that our results are consistent with the predictions based on the study of magnetic CFI in rotating fluids. The results are discussed in the context of AGN jets in general and the nature of the Fanaroff-Riley morphological division of extragalactic radio sources in particular.    
\end{abstract}

\section{Introduction}

Jets from black holes of active galactic nuclei and young stars exhibit remarkable ability to propagate over very large distances, up to $10^9$ of their initial radius at the jet ``engine'' \citep[e.g.][]{Porth15}. This is in contrast to the expectations based on the linear stability analysis of cylindrical jets and laboratory experiments. Cylindrical jets are subject to Kelvin-Helmholtz (KHI) and current-driven instabilities (CDI), for magnetised flows. The e-folding length scale for the fastest growing body modes of KHI is $l\sub{KHI}\approx M_s R_j$, where $M_s$ is the Mach number based on the sound speed and $R_j$ is the initial jet radius \citep[e.g.][]{Hardee87A,Hardee87}.    For CDI the corresponding length scale is $l\sub{CDI}\approx 2\pi P M_a$, where $M_a$ is the Mach number based on the Alfv\'en speed and $P=R_j B\sub{cm}^z/B\sub{cm}^\phi$  is the magnetic pitch as measured in the comoving frame;  $B\sub{cm}^z$ and $B\sub{cm}^{\phi}$ are the comoving axial and azimuthal components of the magnetic field respectively \citep[e.g.][]{ALB00}. Both expressions apply to both Newtonian and relativistic flows, provided one uses the relativistic definitions of the Mach numbers in the latter case.    

One obvious difference between the cylindrical jet models and the real astrophysical jets is that the radius of the real jets is not constant but grows significantly with the distance from their central engine.  This radial expansion of astrophysical jets can be a response to the reduction of external pressure with the distance from their engines. In the extreme case where the external pressure is too low for confinement, jets expand freely and their shape becomes conical. 

This complicates the problem and does not allow a straightforward application of the results obtained in the stability studies of cylindrical jets. For example, which radius should we use in estimating the e-folding length scale?  If we opted for the radius near the jet engine we would obtain a scale which is many orders of magnitude below the observed length of astrophysical jets for any realistic values of the jet parameters.  This would be in conflict with the very existence of these jets. If we opted for the jet radius near its termination, we would get amuch larger e-folding scale, making the stability issue much less severe.     

In fact, the jet radius is important for the development of the body mode KH and CD instabilities  because they rely on communication across the whole jet in order to get amplified. Since the communication speed is limited either by the sound speed or the Alfv\'en speed, an increase of the jet radius implies an increase of the jet crossing time-scale, and hence an increase of the instability growth length-scale. This is why in the case of cylindrical jets these scales are proportional to the jet radius.

\citet{Hardee87} modelled the spacial growth of KHI in expanding jets confined by external pressure $P \propto z^{-a}$ assuming that the local growth rate is the same as in the cylindrical jet with the same flow parameters. He found a significant reduction of the overall growth. Moreover, for $a = 2$ the perturbation amplitude is no longer an exponential function of the distance but a power-law and for $a > 2$ its growth quickly comes to halt. In fact, for $a > 2$ initially pressure-matched jets eventually become free-expanding as sound waves can no longer provide causal communication across the jet \citep{Lyub92}.       

Using the causality argument, \citet{Porth15} concluded that $a=2$ is critical for relativistic magnetised jets as well.  In order to test this conclusion, they used a 3D periodic box setup to study the stability of expanding relativistic jets with the magnetisation  parameter $\sigma\lesssim 1$  and predominantly  azimuthal magnetic field ($\sigma=b^2/4\pi w$, where $b$ is the magnetic field strength in the fluid frame and $w$ is the relativistic enthalpy). Within the box, their jets had cylindrical geometry and their expansion was promoted via a forced decline of the external gas pressure. The temporal rate of the decline was set to what would be seen in a reference frame moving with relativistic speed through the atmosphere with the gas pressure $P\propto z^{-a}$. The results showed progressively increasing reduction of the instability growth rate with increasing value of $a$, leading to its almost complete suppression for $a\geq 2$.  A number of other computation studies support the reduction of the instability growth rates in expanding jets \citep[e.g.][]{RH00,MollSpruit2008,BM-09,Porth2013}. 

The large-scale radio sources created by AGN jets are roughly divided into two classes, luminous FR-2 sources with bright outer regions  and under-luminous FR-1 sources with bright inner regions \citep{FanaroffRiley1974,OL-94}.  These inner regions of FR-sources are dominated by conspicuous and rather broad jets \citep[e.g.][]{RPFF90}. In most cases, these jets seem to enter the extended radio lobes and disappear inside them, and in other cases the radio lobes appear as a continuation of the jets \citep{RPFF90}.  The latter case includes the ``prototype'' FR-1 radio source 3C$\,$31.  Jets of FR-2 sources are much fainter relative to their radio lobes then FR-1 jets. When they are visible, they can often be traced all the way from the central core to the outer hotspots \citep[e.g.][]{LBD-97}. This includes the prototype FR-2 source Cyg A \citep{CB-96}.

Beginning from the pioneering works by \citet{BR-74} and \citet{Sch74}, the large-scale structure of FR-2 sources is explained in terms of the shock interaction between a super-fast-magnetosonic jet and external gas. The model assumes that these jets remain largely intact, and hence untouched by global instabilities, until they reach the impact locations identified with the outer hot spots.   In contrast, the observed morphology of FR-1 jets suggests turbulent dynamics with enhanced internal friction, mixing, and mass entrainment \citep[e.g.][]{bicknell-84,Kom-90}. Since turbulence is a common non-linear outcome of hydrodynamic and magnetohydrodynamic instabilities, it seems reasonable to explore if  these instabilities could be behind the   Fanaroff-Riley division.

The rapid expansion of astrophysical jets can halt when they enter regions with sufficiently high and slowly varying external pressure. This is because the internal pressure of expanding jets decreases very rapidly and may quickly drop below the external pressure. In this case, the external pressure drives a shock, often called a reconfinement shock, into the jet. This shock reheats the jet and  establishes approximate pressure balance with the external gas. Steady-state two-dimensional models of reconfined jets predict that they become  approximately cylindrical, though with quite strong superimposed oscillations. This creates favourable conditions for development of the instabilities, which were previously suppressed in the expansion zone.  

For the jets confined by the interstellar gas (the so-called ``naked'' jets) such conditions can be met inside the central cores of the X-ray coronas of their parent galaxies.  \citet{Porth15} have found that the only FR-1 can be reconfined inside these cores, whereas the reconfinent point of FR-2 jets is located on the distances comparable with the size of their extended lobes.  The cocoons (lobes) of shocked plasma inflated by the jets is another place where only slow variation of pressure is expected.  This is because the expansion speed of these cocoons can be highly subsonic \citep[e.g.][]{Falle1991}. 
     
\citet{Falle1991} proposed a self-similar model for the evolution of the large-scale structures created by FR-2 jets. This model predicts a relatively slow decrease of the cocoon pressure and hence a gradual increase of the jet length to its radius with time (or the source size). \citet{Falle1991} argued that as the length to radius ratio grows sufficiently large,  the jet develops instabilities and become turbulent, and that this results in a transition to the FR-1 morphology. This explanation is supported by the observations showing FR-1 jets broadening  and disappearing inside the radio lobes \citep{RPFF90}. However, these jets may develop instabilities and turbulence closer to the central source, where they may still be naked.  Recent 3D hydrodynamic (HD) and magnetohydrodynamic (MHD) simulations of non-relativistic cylindrical jets by \cite{MassagliaBodo2016,Massa19} also provide some support to this idea.  However in these simulations, the jets are injected into the computational domain as already perfectly collimated flows, bypassing the initial phase of free or almost free expansion. 

\cite{TchekhovskoyBromberg2016} carried out relativistic 3D MHD simulations of outflows generated by a rotating sphere with monopole magnetic field. These outflows also inflated cocoons of hot gas which provided their confinement and quasi-cylindrical collimation. These cylindrical flows suffered from CDI kink modes, which in some cases led to a development of morphology reminiscent of FR-1 radio sources. However, these were not proper jets but rather the so-called ``magnetic towers'' \citep{Lynden-Bell-03}, as the flow  speed remained sub-fast magnetosonic. 

The recollimation of supersonic (super-fast-magnetosonic in the magnetic case) jets may be accompanied by another instability (which we tentatively call the {\it recollimation instability}), which is not present in cylindrical configurations. This possibility was first recognised by \citet{MatsumotoMasada2013}, who  argued that the accelerated transverse motion associated with the radial oscillations of reconfined jets  is similar to the radial oscillations of non-equilibrium cylindrical jets. Hence they have shown that the oscillations of cylindrical jets are accompanied by the Rayleigh-Taylor instability \citep[RTI,][]{Rayleigh:1883, Taylor:1950} at the interface between the jet and external medium. They explored this via 2D relativistic HD simulations of oscillating cylindrical jets, which have also demonstrated that the overall effect can be amplified via the Richtmyer-Meshkov instability \citep[RMI,][]{Richtmyer:1960,Meshkov:1972} of the shocks associated with the oscillations. \citet{Matsumoto17} carried the linear stability analysis of the relativistic RTI using the incompressibility approximation.  

Recently, several groups carried out 2D and 3D simulations of non-magnetic reconfined jets \citep{Gourgouliatos18a,Gottlieb19,Gottlieb20c,Gottlieb-20a,MM19}.   They have demonstrated that the recollimation instability develops only in 3D simulations, where it can lead to a rapid transition to a fully turbulent state soon downstream the reconfinement point, in great contrast to the predictions of steady-state axisymmetric models.  \citet{Gourgouliatos18a, GK18} argued that the recollimation instability is related not to RTI but to the centrifugal instability \citep[CFI,][]{Rayleigh:1917}.  The relativistic version of this instability in rotating flows was studied by \citet{GK18}.  

AGN jets carry out magnetic field, and it is known that a sufficiently strong magnetic field can   inhibit various hydrodynamic instabilities when their development lead to an increase of the field energy.  In particular, \citet{KGM19} studied the role of axial magnetic field on the development of CFI at the cylindrical interface between rotating relativistic fluids.  Extrapolating the results to the problem of reconfined jets, they concluded that a relatively weak magnetic field, with $\sigma=0.01-0.1$ may completely suppress the recollimation instability in reconfined jets. Here we investigate this problem directly, via 3D relativistic MHD simulations of AGN jets. 

The structure of the paper is as follows. In section \ref{METHOD}, we present our method and the setup of the simulations we performed. Section \ref{RESULTS} describes the results of these simulations. We discuss the astrophysical implications of these results in section \ref{DISCUSSION} and summarise our conclusions in section \ref{CONCLUSION}.

\section{Method}
\label{METHOD}

\subsection{Overview} 

In this study, we use computer simulations to investigate the role of the magnetic field on the stability of relativistic jets undergoing reconfinement by the thermal pressure of external gas. In order to ensure the continuity with the previous studies and allow for direct comparison, we use as a starting point the non-magnetic model C1 of \citet{Gourgouliatos18a}. In this model, an initially conical jet propagates through the X-ray corona of the parent galaxy.  The pressure distribution of the corona is modelled  with the isothermal King law.  The initial solution describes a steady-state jet in direct contact with the external gas. There is no cocoon separating the jet from the external gas. This configuration corresponds to the sub-group of FR-1 jets whose morphology is similar to the jets of the radio source 3C31  (the so-called naked jets).  

The C1 jet exhibits rapid development of the recollimation instability and hence this model provides a good reference for studying the role of magnetic field in this process. To this aim, we modify the C1 model by adding a purely azimuthal magnetic field, while keeping other parameters unchanged.  On the scale of galactic coronae, the azimuthal component is expected to dominate over the poloidal magnetic field emerging from the jet engine. The polarisation observations of FR-1 jets also indicate the presence of a longitudinal component. However it is normally attributed to the small scale irregular magnetic field \citep[e.g.][]{Laing-81,BBR-84,Wardle-13}, which is unlikely to influence long wavelength instabilities.         
  
The overall strategy of our computational experiments is the same as  in \citet{Gourgouliatos18a}.  First, we find an approximate axisymmetric steady-state solution using the method described in \cite{Matsumoto12, Komissarov15}.  Next, we use the result to setup initial conditions for time-dependent axisymmetric simulations. 
The purpose of these 2D simulations is 1) to check that the steady-state solutions are sufficiently accurate and 2) to see if they develop axisymmetric instabilities. Finally, we use the steady-state solution to setup initial conditions for fully three-dimensional simulations.

\subsection{Governing equations}
We solve the equation of ideal special relativistic MHD. These are the continuity equation 
\beq
\partial_{\alpha}(\rho u^{\alpha})=0\;, \label{eq: mass conservation}
\eeq
the energy-momentum equation
\beq
\partial_{\alpha} \biggl[ \left(w + \frac{b^2}{4\pi}\right)u^{\alpha}u^{\beta} - \frac{b^{\alpha}b^{\beta}}{4\pi} + \biggl ( P + \frac{b^2}{8\pi} \biggr ) g^{\alpha \beta} \biggr ]=0\;, \label{eq: energy-momentum conservation}
\eeq
and the Faraday equation
\beq
\partial_{\alpha}(u^{\alpha}b^{\beta} - u^{\beta}b^{\alpha})=0\;. \label{eq: Maxwell}
\eeq
Here $\rho$ is the rest-mass density, $u^{\alpha}$ is the four-velocity, $w=\rho c^2+\gamma/(\gamma-1)P$ is the relativistic enthalpy of ideal gas, $P$ is the gas pressure, $b^{\alpha}$ is the 4-vector of magnetic field, and $g^{\alpha \beta}$ is the metric tensor of Minkowski spacetime. In the simulations we use the ratio of specific heats $\gamma = 4/3$. 

The 3+1 decomposition of $u^\alpha$ and $b^{\alpha}$ is 
\begin{align}
u^\alpha &=\Gamma (c, \vv{v})\;,\\
b^{\alpha}  &=\Gamma \biggl ( (\spr{v}{B}), \; \frac{\vv{B}}{\Gamma^2} + \vv{v}(\spr{v}{B}) \biggr ) \;, \label{eq: b-field}
\end{align}
where, $\Gamma$ is the Lorentz factor and $\vv{v}$ is the 3-velocity, and \vv{B} is the magnetic field vector as measured in the laboratory frame. The strength of the magnetic field in the fluid frame is $B\sub{cm}=b$, where
\beq
b^2 =b_\alpha b^\alpha =\frac{B^2}{\Gamma^2} + (\spr{v}{B})^2\;, \label{eq: b2}
\eeq

In order to simplify identification of the jet plasma we introduce a passive tracer, $\tau$,  governed by the equation
\beq
\partial_{\alpha}(\rho \tau u^{\alpha})=0\;. \label{eq: tracer}
\eeq
In the initial solution, $\tau$ is set to unity inside the jet and to zero in the external medium. Moreover, it is kept at unity in the ghost cells of the nozzle boundary, so that the injected flow carries this value of the tracer into the computational domain.

The equations are integrated using the AMR-VAC code as described in \citet{Keppens12} and \citet{Porth14}. The relativistic HLLC scheme, Koren flux limiter and two-step Runge-Kutta for time integration are selected for the simulations. 

\begin{figure}
\begin{center}
\includegraphics[width=\columnwidth]{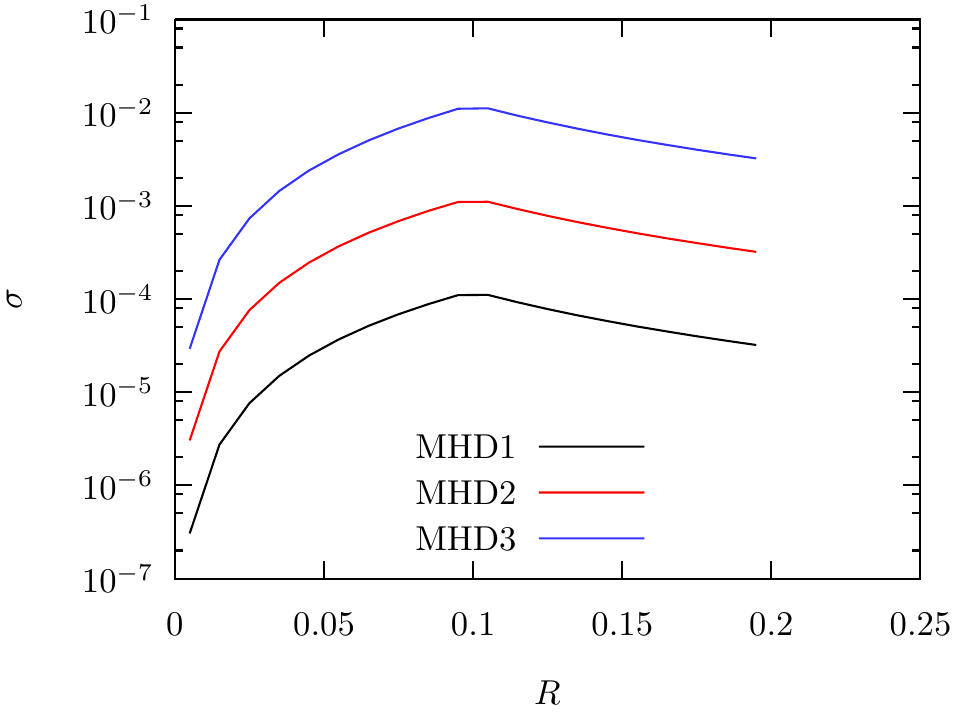}
\caption{Distribution of the magnetisation parameter $\sigma$ at the jet nozzle.}
\label{fig:sigma-nozzle}
\end{center}
\end{figure}

\subsection{Jet setup} 

The external gas is assumed to be isothermal, with a spherically-symmetric mass density distribution described by the King law 
\beq
\rho_{e}(r) = \rho_{e,0} \biggl ( 1 + \frac{r^2}{r_c^2} \biggr ) ^{-a/2} \,, \label{external density distribution}
\eeq
where $r$ is the spherical radial coordinate,  $r_c$ is the core radius, and $\rho_{e,0}$ is the central density.
The power-law index of the density distribution is set to $a=1.25$, the typical value for giant elliptical galaxies. 

The jet nozzle is located at the distance $z_0=0.1 r_c$ from the origin, with the initial radius $R_0=0.02 r_c$.  The jet density distribution at the nozzle is uniform $\rho=\rho_{j,0}$. Initially, the jet is relativistically cold, with $P\ll \rho c^2$ and hence $w\approx \rho c^2$.  

The velocity distribution over the nozzle corresponds to a conical flow of the half-opening angle $\theta_0=0.2$ emerging from the origin:  
\beq
(v_R, \; v_{\phi}, \; v_z) = v({\rm sin} \; \theta, \; 0, \; {\rm cos} \; \theta) \; ,
\eeq
where $\{R,\phi,z\}$ are cylindrical coordinates aligned with the jet axis and $\theta=$ arctan $(R/z_0)$. The corresponding Lorentz factor depends only the cylindrical radius, 

\beq
\Gamma= \Gamma_0 \biggl [ 1- \biggl ( \frac{R}{R_0} \biggr )^4 \biggr ]+ \biggl ( \frac{R}{R_0} \biggr )^4 \;. 
\eeq
This is a smooth approximation of the top-hat profile with $\Gamma=\Gamma_0$ at $R=0$ and $\Gamma=1$ at $R=R_0$. In the simulations, we used $\Gamma_0=5$. 

The magnetic field is assumed to be purely azimuthal. At the nozzle it has the core-envelope distribution
\beq
b^{\phi}(R) = \left\{ 
    \begin{array}{ccl}
         b^{\phi}_m (R/R_m) & \mbox{if} & R < R_m \;, \\
         b^{\phi}_m (R_m/R) & \mbox{if} & R_m \leq R \leq R_0 \;, \\
         0 & \mbox{if} & R_0 < R \;,
    \end{array} \right.
\eeq
where $R_m$ is the core radius.   The electric current is uniform inside the core and vanishes in the envelope. The return current flows over the jet surface $R=R_0$.  All our models have $R_m=R_0/2$.

The relativistic magnetisation parameter $\sigma$ is maximum at the magnetic core radius, where it reaches the value
\beq
\sigma\sub{max} = \frac{{b_m^\phi}^2}{4 \pi w} \;.
\eeq
We have studied four models with $\sigma\sub{max}=0$ (HD), $10^{-4}$ (MHD1), $10^{-3}$ (MHD2), and $10^{-2}$ (MHD3).  This range was found to be sufficient for the purpose of the study. Figure \ref{fig:sigma-nozzle} shows the radial distribution of $\sigma$ at the nozzle.  
 
Table~\ref{table1} summarises the physical parameters of our numerical models. The dimensional parameters are scaled to yield  $c=3\times 10^{10} \mbox{cm}\, \mbox{s}^{-1}$, $r_c=1\,$kpc, and the jet power $L=2 \times 10^{44} \mbox{erg}\, \mbox{s}^{-1}$. The latter roughly corresponds to the boundary between FR-1 and FR-2 sources.
 
\begin{table}
\caption{Physical parameters of simulations}
\begin{tabular}{lc}
\hline
 & \\
Coronal core radius  & $r_c=1\,$kpc\\
Coronal core density & $\rho_{e,0}=3.3\times10^{-27}\mbox{g}\,\mbox{cm}^{-3}$ \\
Coronal core pressure & $P_{e,0} = 3\times 10^{-10}\mbox{dyn}\,\mbox{cm}^{-2}$ \\
Jet nozzle position & $z_0=0.1 r_c$\\
Jet nozzle radius & $R_0=0.02 r_c$ \\
Initial jet half-opening angle & $\theta_0=0.2$ \\
Initial jet Lorentz factor & $\Gamma_0=5$\\
Initial jet density & $\rho_{j,0}=5.3\times10^{-29}\mbox{g}\,\mbox{cm}^{-3}$ \\
Initial jet magnetization & $\sigma_{max} = 0$ (HD),  $10^{-4}$ (MHD1),\\
&   $10^{-3}$ (MHD2), $10^{-2}$ (MHD3) \\
Jet power & $L=2\times 10^{44} \mbox{erg}\,\mbox{s}^{-1}$ \\
 & \\
\hline
\label{table1}
\end{tabular}
\end{table}

\begin{figure*}
\begin{center}
\scalebox{0.4}{{\includegraphics{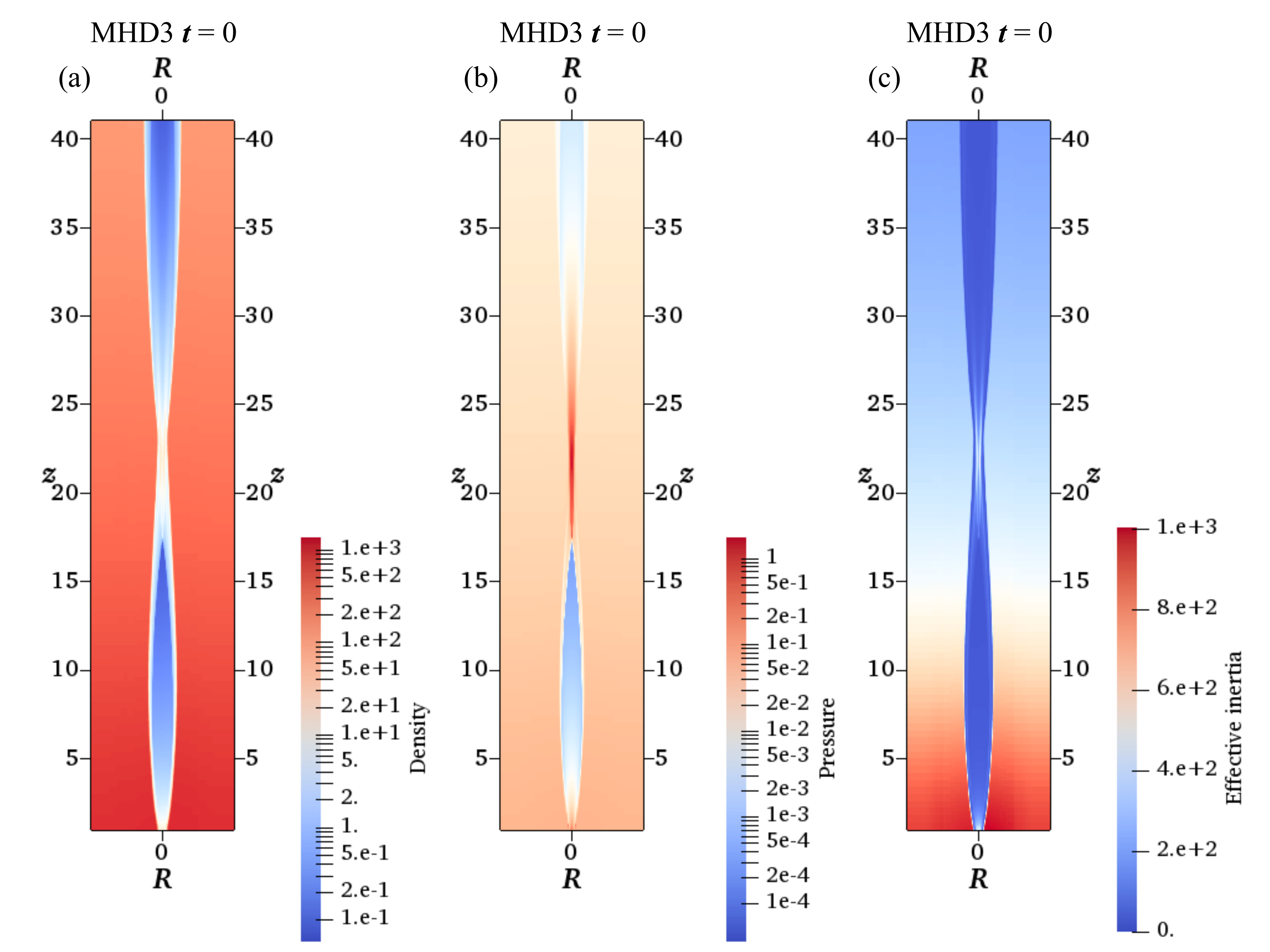}}}
\scalebox{0.4}{{\includegraphics{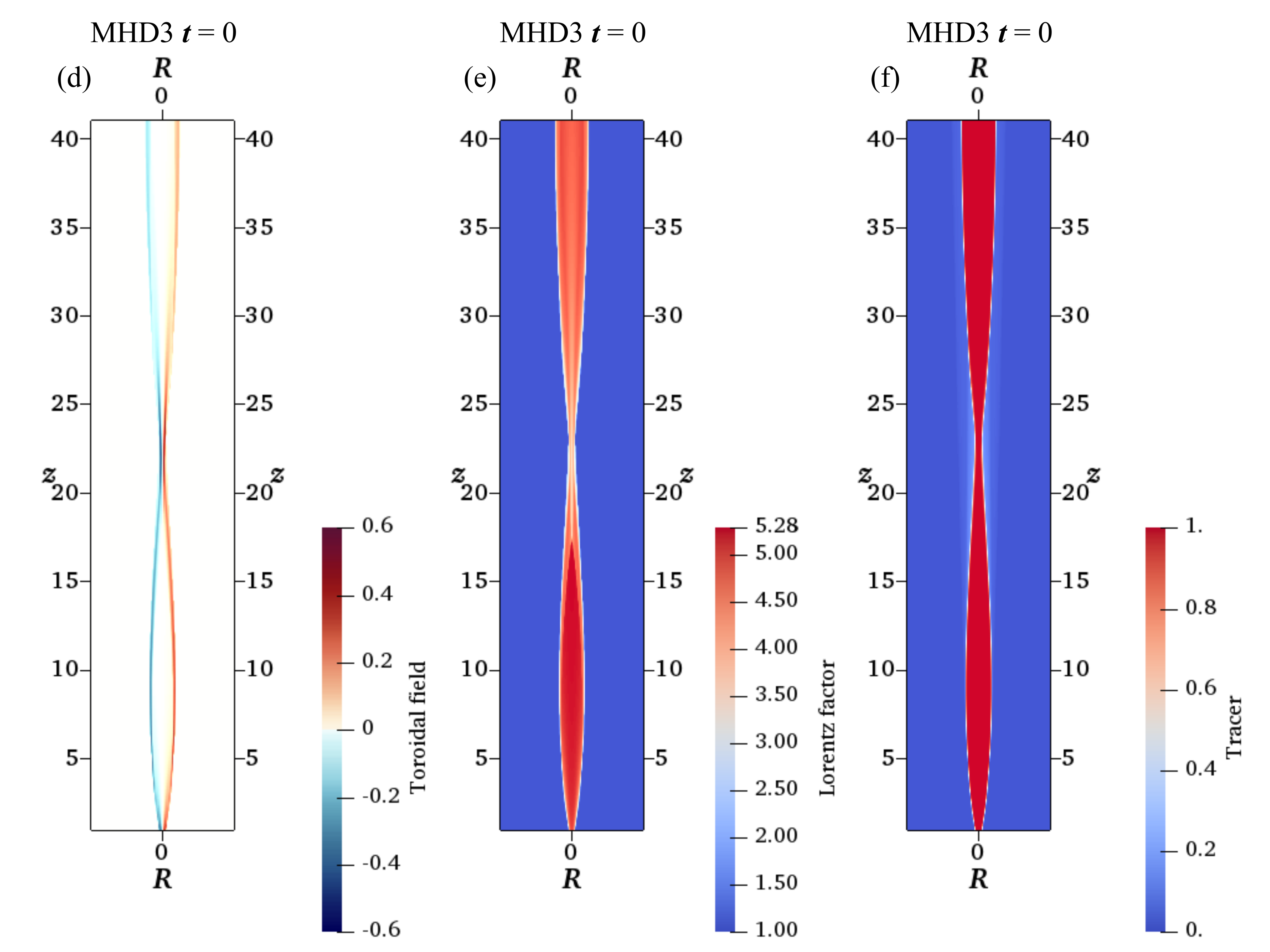}}}
\caption{Steady-state solution of the model MHD3.  The plots show the distributions of the density $\rho$ (panel a), gas pressure $P$ (panel b), effective gas inertia $w \Gamma^2 $ (panel c), magnetic field $b^{\phi}$ (panel d), Lorentz factor $\Gamma$ (panel e), and tracer $\tau$ (panel f).}
\label{fig:steady-state}
\end{center}
\end{figure*}

\subsection{Steady-state solutions}

The complex steady-state solutions describing reconfined jets were constructed using one-dimensional simulations of flows with cylindrical symmetry, following the technique developed in \citep{Matsumoto12, Komissarov15}.  Standard time-dependent two-dimensional axisymmetric simulations were used to verify their suitability as initial solutions for the instability study.

\begin{figure*}
\begin{center}
\scalebox{0.9}{{\includegraphics{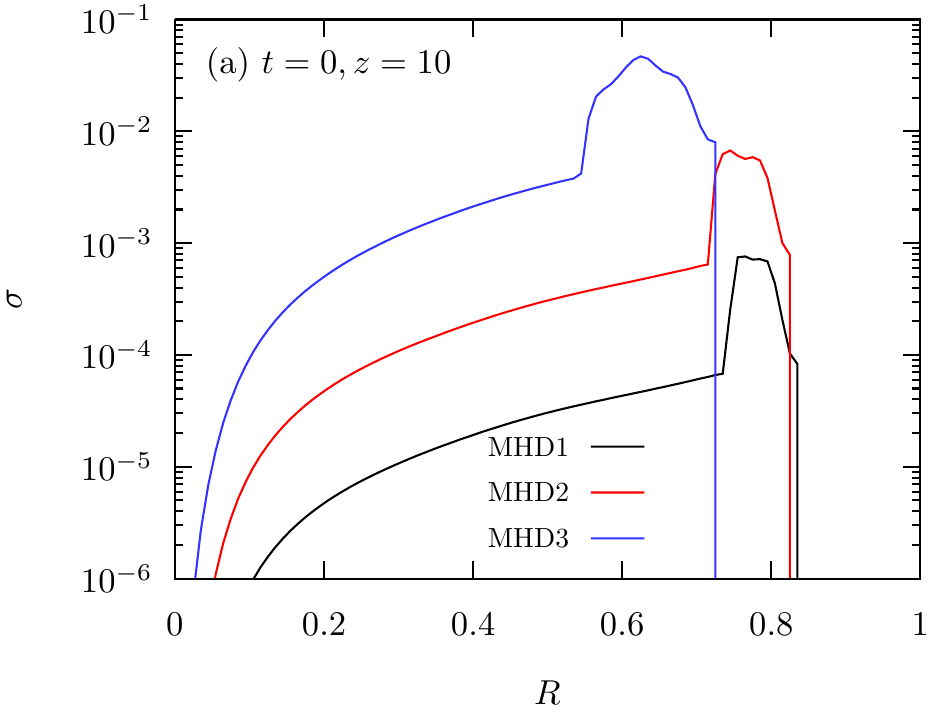}}} 
\scalebox{0.9}{{\includegraphics{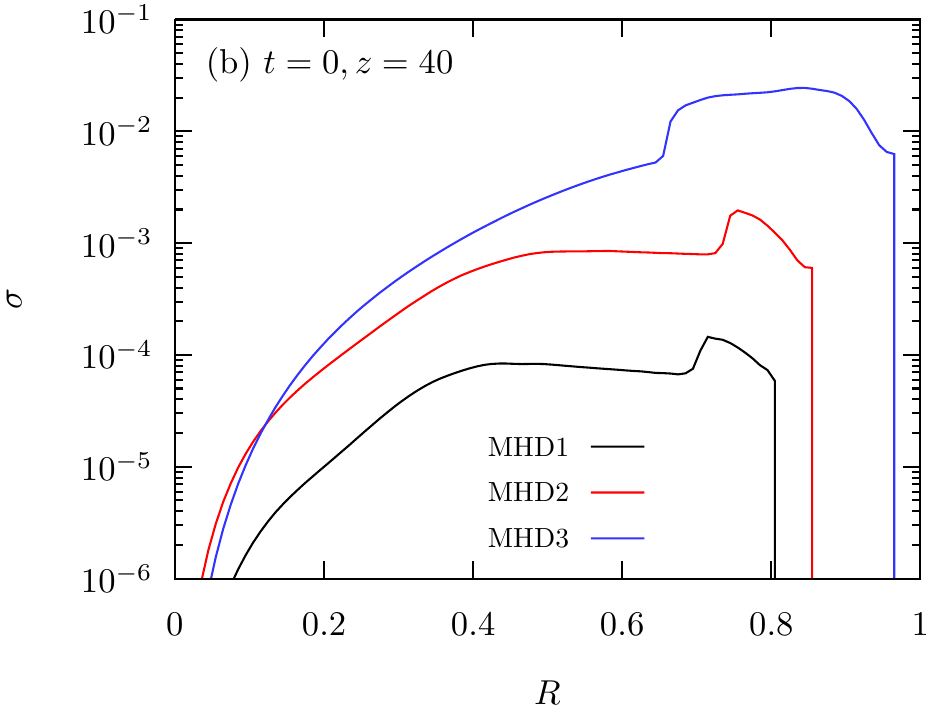}}} 
\caption{Radial distribution of the magnetisation parameter $\sigma$ for the steady-state solutions at $z=10$ and $40$.}
\label{fig:sigma}
\end{center}
\end{figure*}

\subsubsection{``One-dimensional'' models}

In the approach of  \citet{Matsumoto12, Komissarov15}, equilibria solutions of two-dimensional axisymmetric relativistic jet problems are approximated by solutions of time-dependent one-dimensional axisymmetric problems (in the radial direction).  According to this approach, 1) the initial configuration of the time-dependent problem describes the distribution of the flow variables at the nozzle, and 2) the time evolution is triggered via forced variation of the external pressure.
The approximate steady-state solutions $\vv{A}(z,R)$ are obtained from the corresponding time-dependent solution $\bar{\vv{A}}(t,R)$ via the transformation $\vv{A}(z,R)=\bar{\vv{A}}((z-z_0)/c,R)$. For further details see \citet{Komissarov15}.

In our case, the external pressure was varied according to the prescription 
\beq
P_{e}(r(t)) = P_{e,0} \biggl ( 1 + \frac{r(t)^2}{r_c^2} \biggr ) ^{-a/2} \label{external pressure distribution}
\eeq
every time step, where 
\beq
r(t) = \sqrt{R_j(t)^2 + (z_0+ct)^2} \;,
\eeq
where $R_j(t)$ is the jet radius at time $t$.       
For these simulations we used a uniform grid with the cell size $\Delta R = 0.05 R_0$, corresponding to 20 cells per initial jet radius at the nozzle.

Figure~\ref{fig:steady-state} illustrates the properties of the approximate steady-state solutions obtained using this approach, with the model MHD3 used as an example.  Given the relative weakness of the magnetic field,  the global structure of the solutions for other models are not much different, except for the strength of the magnetic field itself.  The most important feature of the solution is the reconfinement shock driven into the jet by the external gas pressure. Initially, both its radius and the jet radius increase but eventually the jet ram pressure drops too low and both radii begin to contract.  At $z\sub{RP}\approx 17$ ( in the units of $z_0=0.1 r_c$) the reconfinement shock converges at the jet axis (the reconfinement point) and gets reflected as a decollimation shock. In the unshocked inner part of the jet, its mass density and magnetic energy density decrease approximately as $z^{-2}$. At the reconfinement shock both these parameters increase and in the shocked outer layer they evolve relatively slowly. As the result, the jet is almost hollow.   The distance to the recollimation point decreases with the jet magnetization. In the HD and MHD1 models it is  $z\sub{RP} \approx 23$, and in the MHD2 model  $z\sub{RP} \approx 22$. The same trend, but at much higher $\sigma$, has been seen by \citet{Fromm-17}.  
 
Figure~\ref{fig:sigma} shows the distribution of $\sigma$ in two cross-sections, one at about half way to the reconfinement point  ($z=10$) and another well downstream of this point and near the far boundary of the computational domain  ($z=40$). One can see that the magnetisation peaks inside the shocked outer layer, where its value exceeds $\sigma\sub{max}$ at the jet nozzle.  This is consisted with the increase of $\sigma$ at fast shocks (reference).     

\subsubsection{Two-dimensional models}

 The initial solutions for the 2D simulations were set via projecting the approximate steady-state solutions (obtained as described in the previous section) on the computational grid of cylindrical coordinates $\{R,z\}$.  Prior to their projection, the density distribution of the steady-state solutions was modified. Following  \citet{Gourgouliatos18a}, its step-like transition between the jet and the external  was replaced with a $tanh$-profile of thickness $\delta R=0.1 R_0$, where $R_0$ is the jet radius at the nozzle. This allowed us to substantially reduce the numerical dissipation at the interface, which otherwise would be too strong and corrupted the solution.

The computational domain was $[0,6]\times[1,41]$, with a uniform grid of $600 \times 200$ cells  (the cell sizes $\Delta R=0.01$ and $\Delta z=0.2$).  This gives  the same radial resolution as in the 1D simulations.  The lower $z$ boundary ($z=1$) was divided into the nozzle section ($0 < R < 0.2$) and the corona section ($R > 0.2$). The values of the physical variables in the ghost cells of the nozzle section were fixed, which is allowed because the jets are super-fast-magnetosonic. In the corona section, we used reflective boundary conditions.  Outflow (zero gradient) boundary conditions were imposed at the $z=41$ and $R=6$ boundaries, and reflective boundary conditions at $R=0$. 

The Newtonian gravity model was used to maintain the hydrostatic equilibrium of the unperturbed coronal gas \citep{Perucho07b}.  This involved an introduction of source terms both in the energy equation and in the momentum equations. These had a little effect on our relativistic jets. 

The 2D simulations were run for 2.5 light-crossing times of the computational domain in the $z$ direction. During this time the solution evolved but its deviation from the initial solution was relatively mild.  There were no signs of instabilities.  This had allowed us to conclude that the initial solutions were close to a steady-state,  stable to axisymmetric perturbations, and hence suitable for use in 3D simulations.  

\begin{figure*}
\begin{center}
\scalebox{0.35}{{\includegraphics{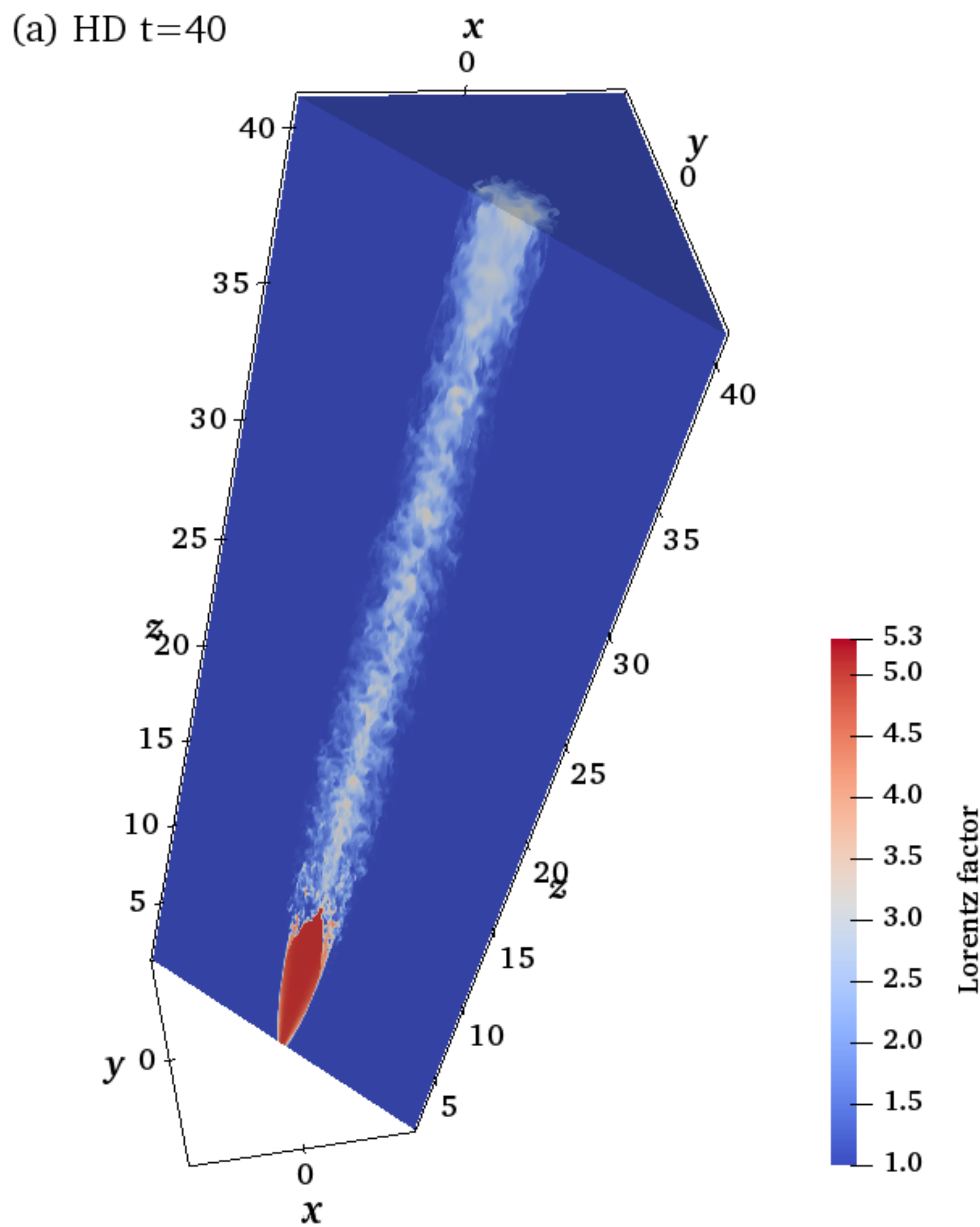}}} 
\scalebox{0.35}{{\includegraphics{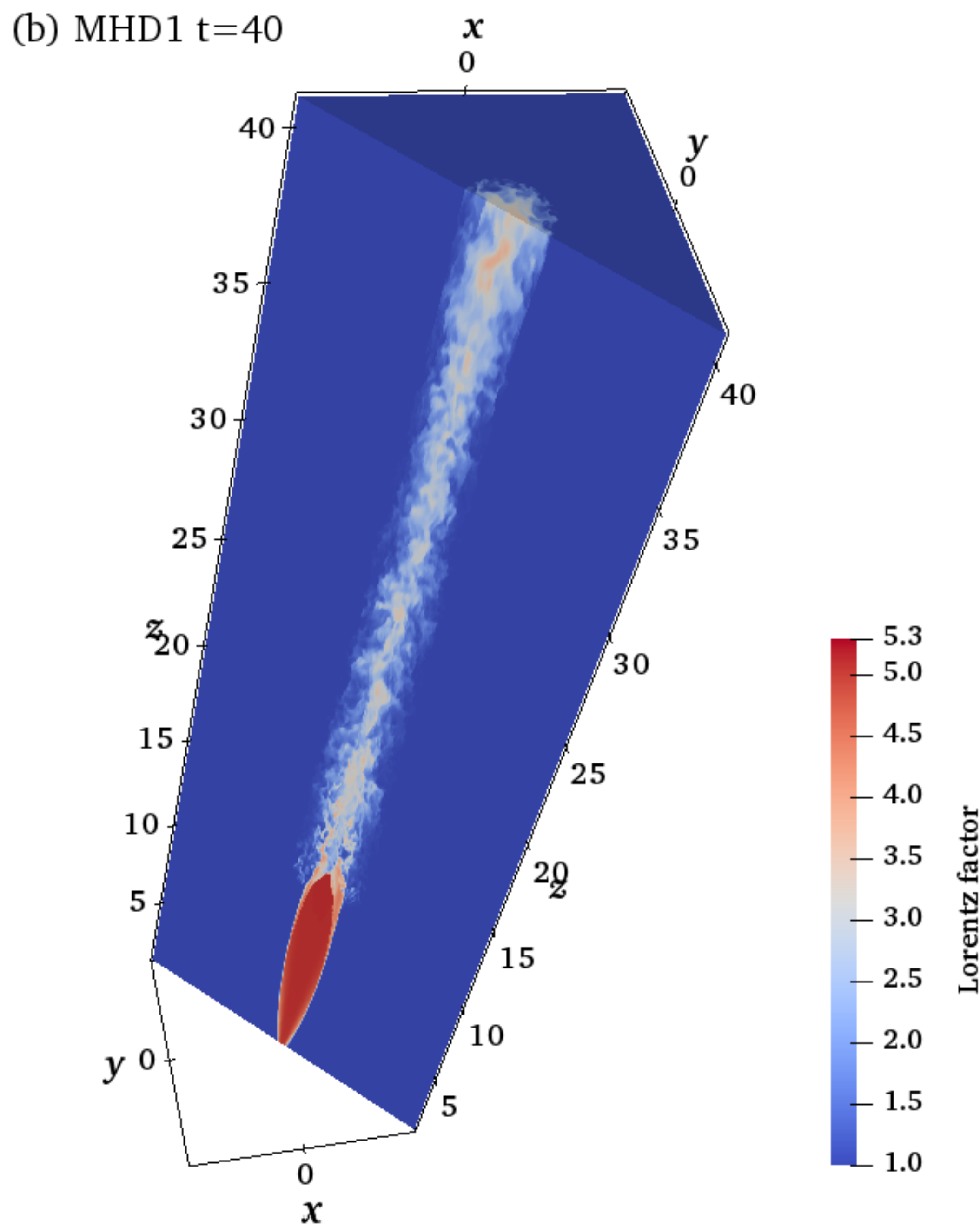}}} 
\scalebox{0.35}{{\includegraphics{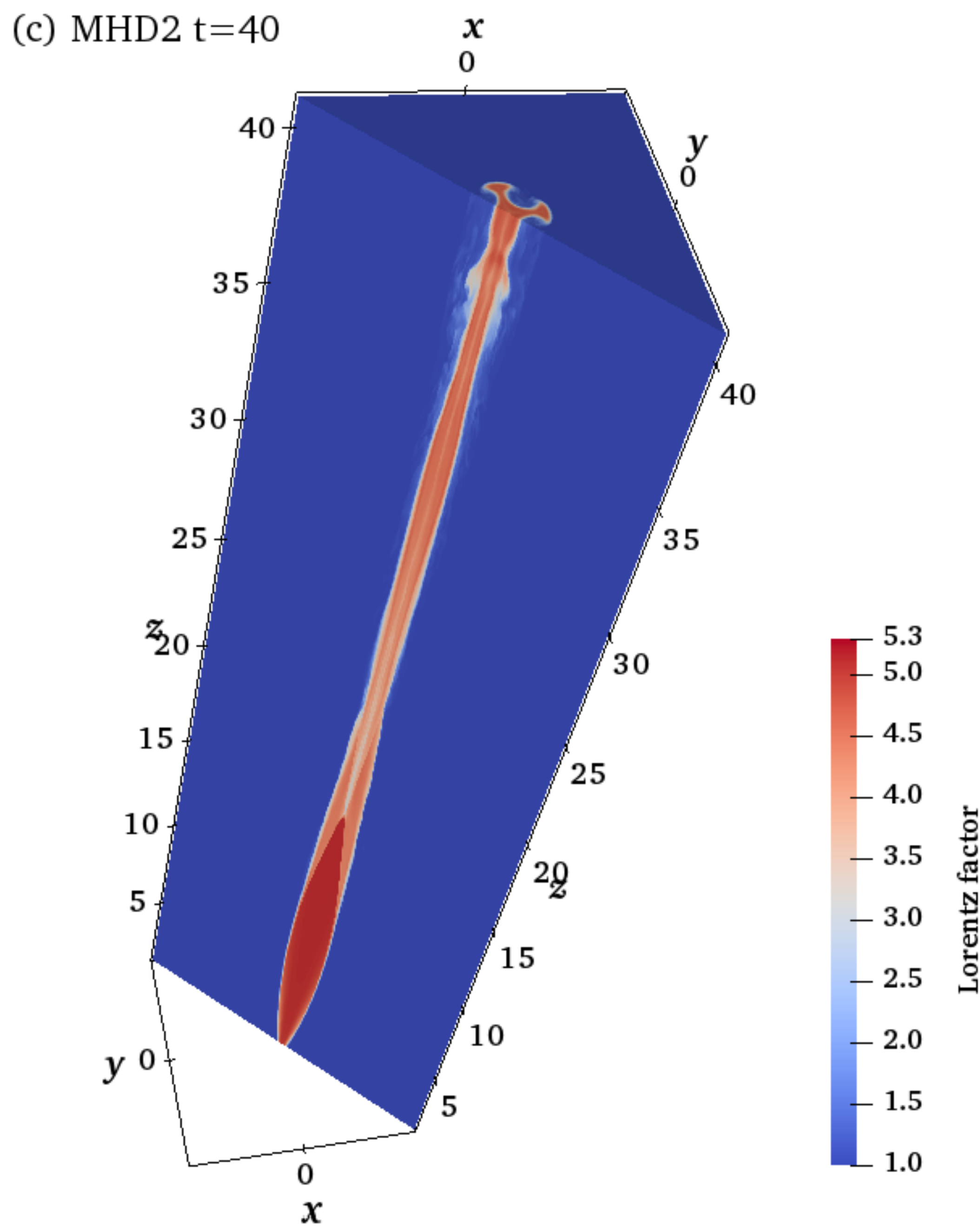}}} 
\scalebox{0.35}{{\includegraphics{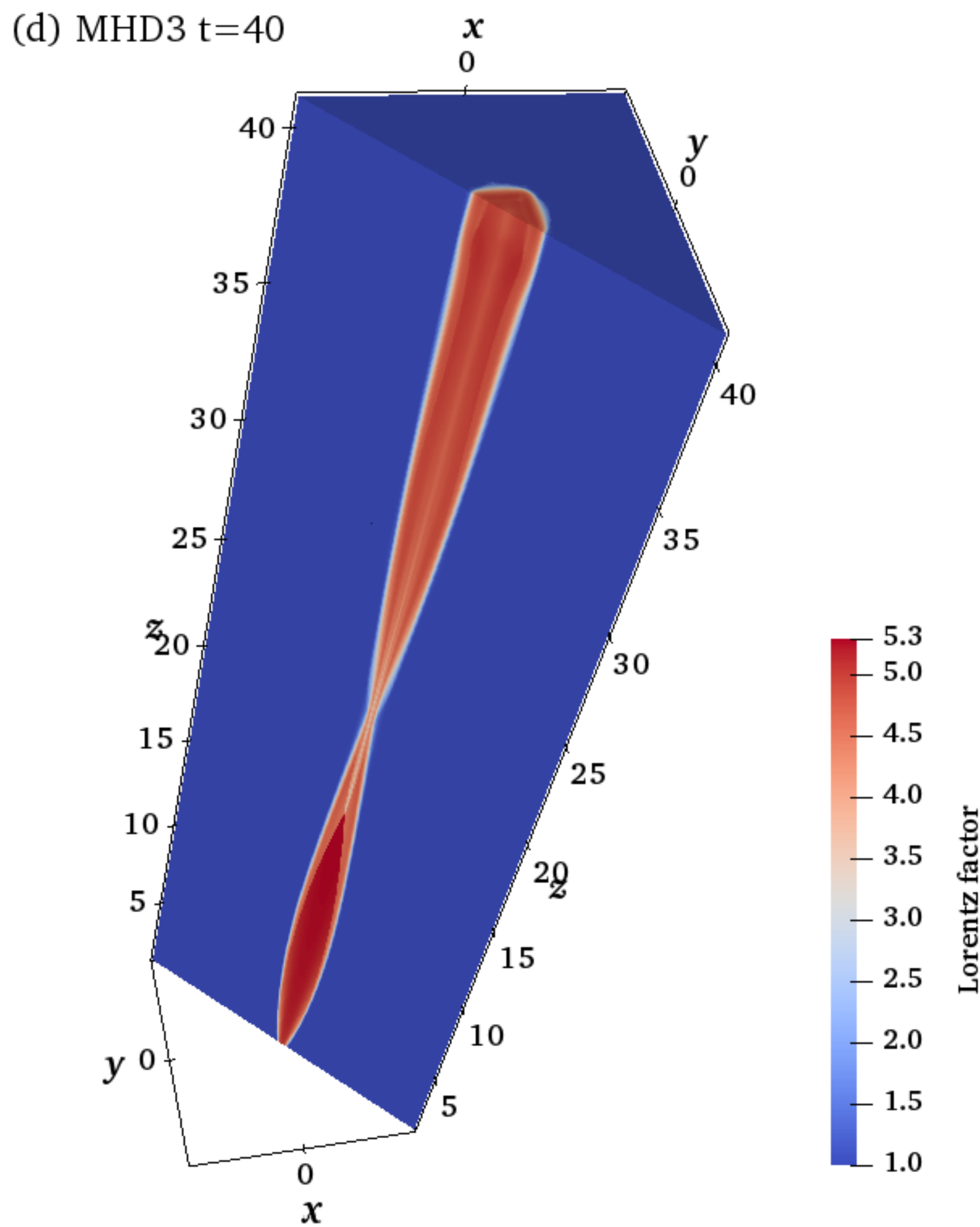}}} 
\caption{3D rendering of the Lorentz factor in 3D solutions at $t=40$. }
\label{3Dgammat40}
\end{center}
\end{figure*}

\subsection{3D time-dependent simulations}

The 3D simulations were carried out on the Cartesian grid of $\{x,y,z\}$ coordinates, with
the computational domain $[-4, 4] \times [-4, 4] \times [1,41]$. In order to reduce the computational cost of the simulations, we capitalised on the adaptive mesh capabilities of the AMRVAC code. We used four levels of adaptive mesh refinement, with  $100^3$ cells at 
the base level. The corresponding cell sizes are $\Delta z =5\Delta x =5 \Delta y $.  At the finest mesh, this was equivalent to the same resolution of 20 cells per nozzle radius at the finest grid as in the auxiliary 1D and 2D simulations.  The refinement was controlled according to the Lohner criterion, with the Lorentz factor as a reference parameter. 

We used the same boundary conditions as in the 2D simulations, slightly adjusted to the different grid geometry (Obviously, the jet axis is no longer a boundary of the simulation domain, and hence no boundary conditions are needed there.). The same applies to the initial setup and the use of Newtonian gravity.   Following \citet{Gourgouliatos18a}, the initial distributions of jet density and pressure were perturbed as 
\begin{eqnarray}
\rho(x,y,z) = \rho_s(x,y,z)(1+10^{-2}\cos\phi) \; , \\
P(x,y,z) = P_s(x,y,z)(1+10^{-2}\sin\phi) \; ,
\end{eqnarray}
where $\rho_{s}(x,y,z)$ and $P_{s}(x,y,z)$ describe the steady-state solution
and $\phi$ is the azimuthal angle.  If such perturbations are not introduced, the growing perturbations are dominated by the mode with the azimuthal number m = 4. This mode is aligned with the Cartresian grid, which is consistent with perturbations arising due to the discretisation errors of the numerical scheme \citep{Gourgouliatos18a}.

\section{Results}
\label{RESULTS}

\begin{figure}
\begin{center}
\scalebox{0.25}{{\includegraphics{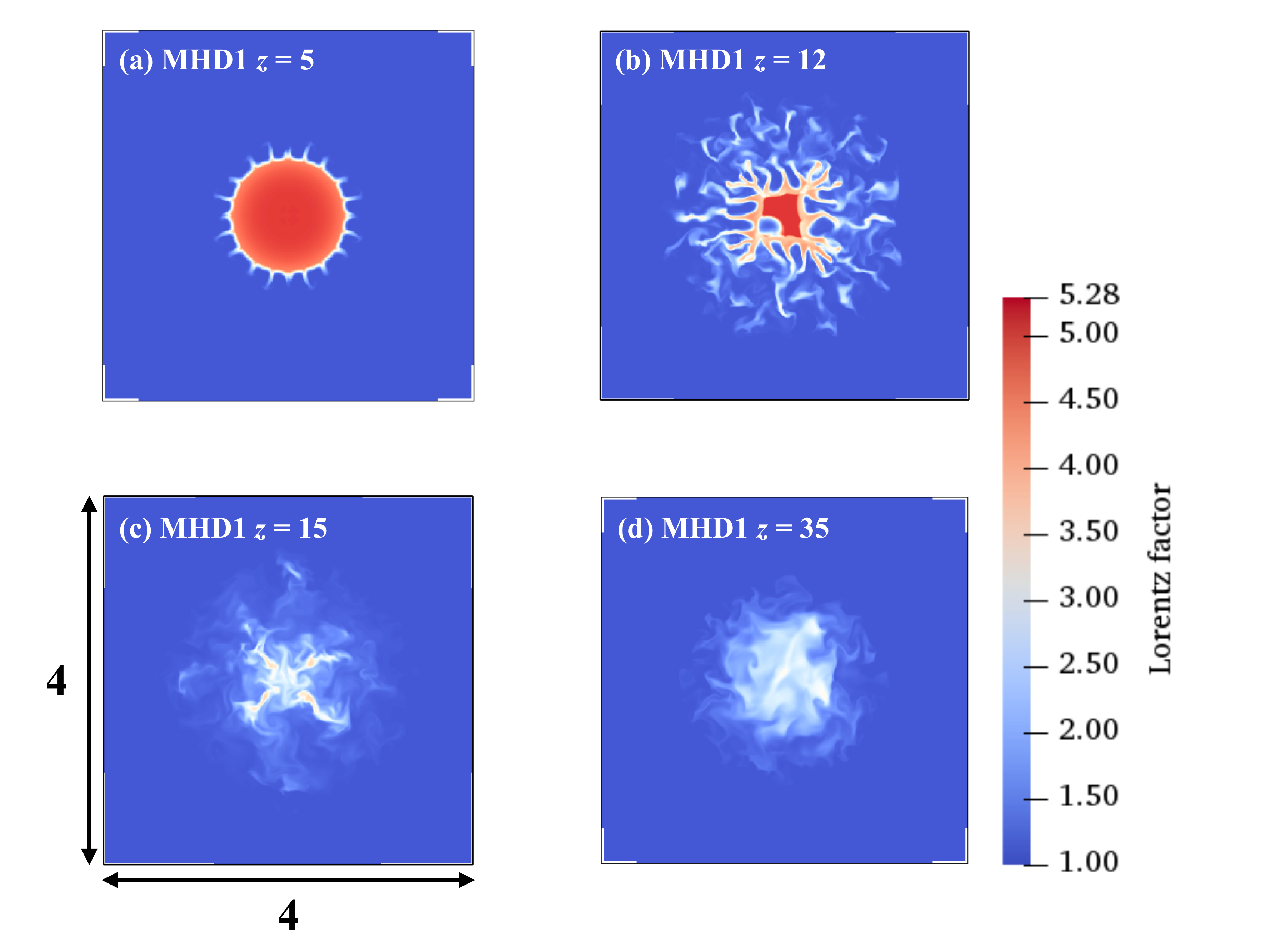}}} 
\caption{Distribution of the Lorentz factor in the model MHD1 at $t=40$ in the cross-sections $z=5,12,15,$ and 35.}
\label{fig:cs-mhd1}
\end{center}
\end{figure}

\begin{figure}
\begin{center}
\scalebox{0.25}{{\includegraphics{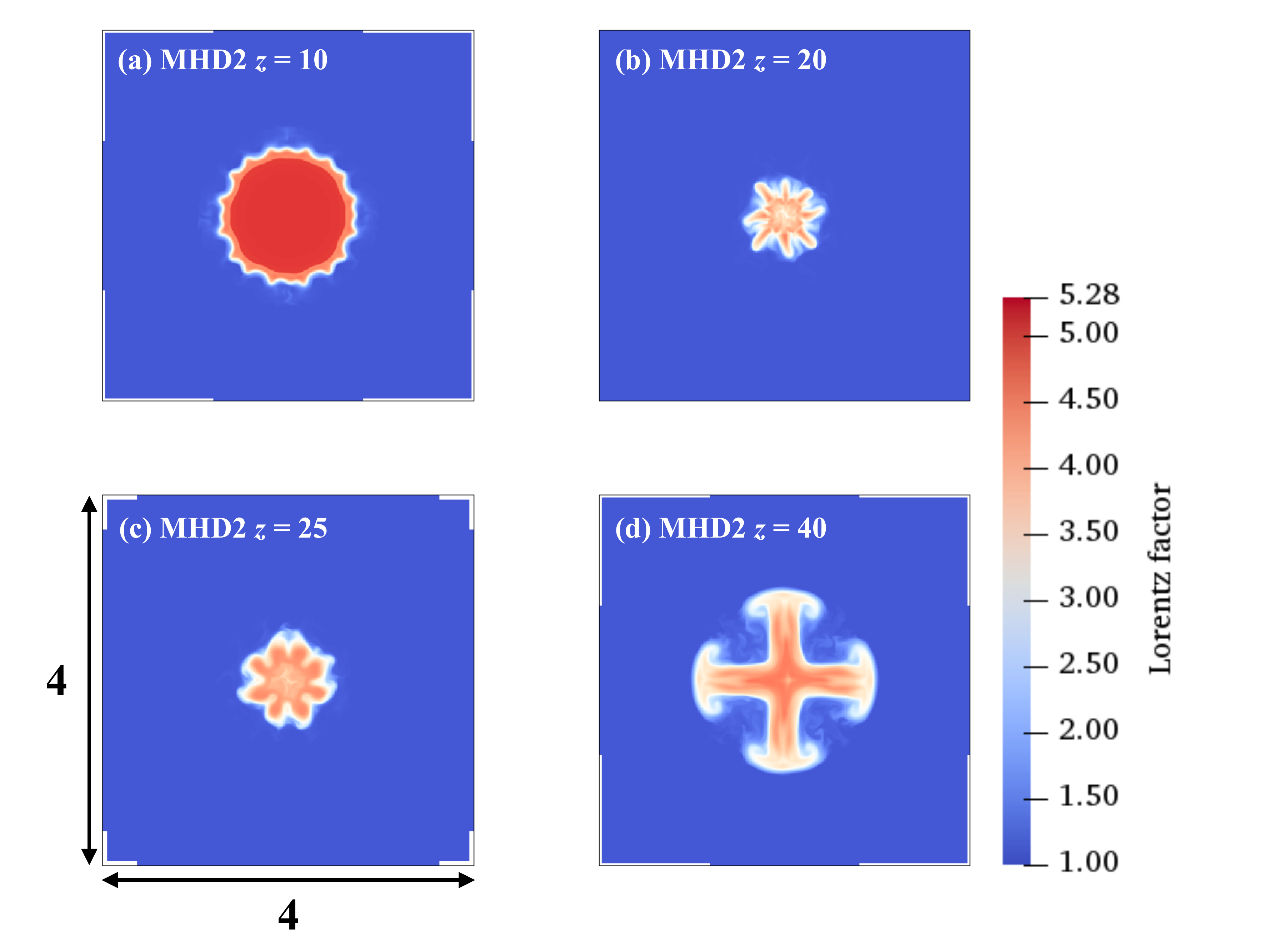}}} 
\caption{The same as in figure \ref{fig:cs-mhd1} but for the model MHD2.}
\label{fig:cs-mhd2}
\end{center}
\end{figure}

\begin{figure}
\begin{center}
\scalebox{0.25}{{\includegraphics{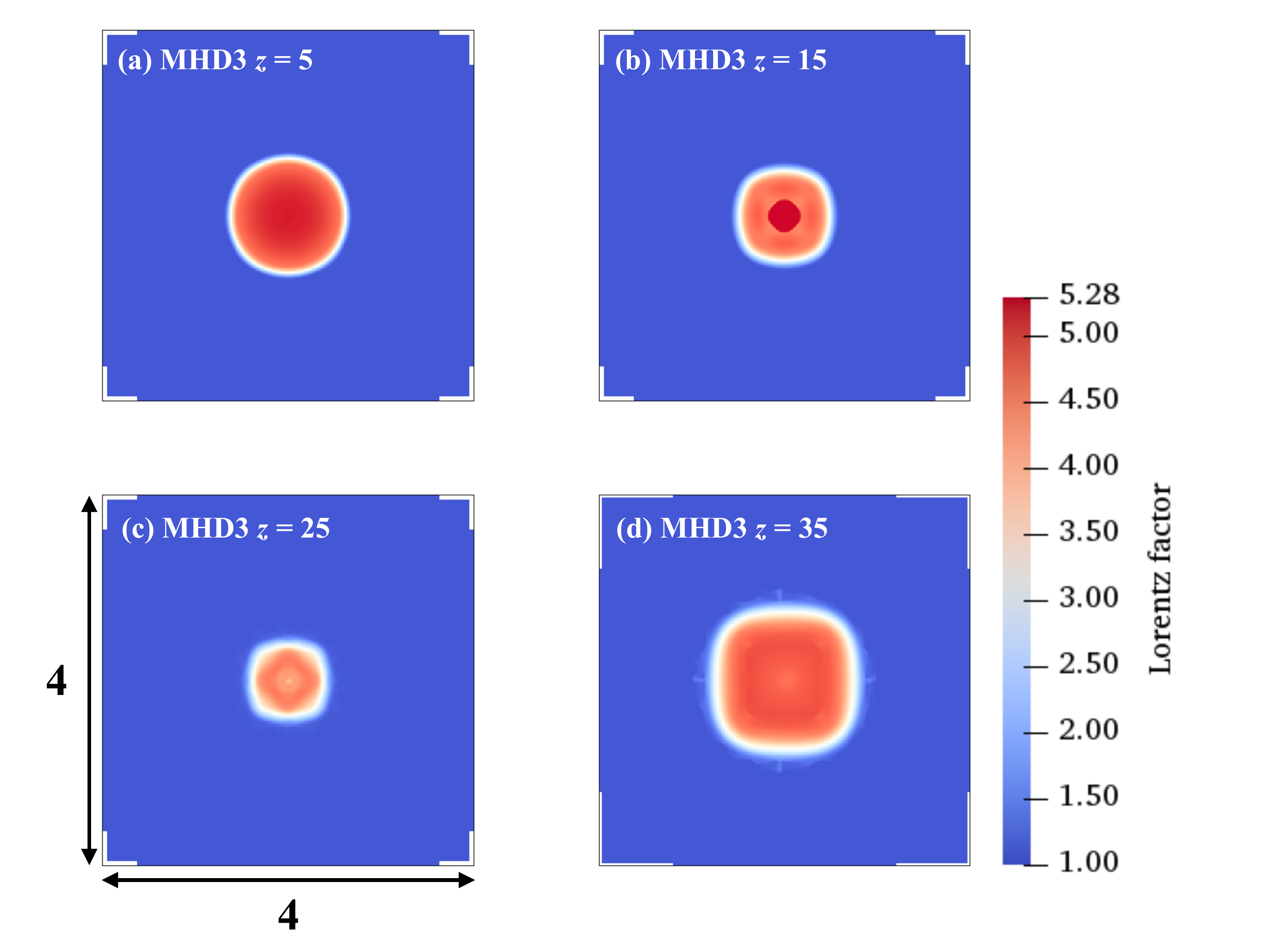}}} 
\caption{The same as in figure \ref{fig:cs-mhd1} but for the model MHD3.}
\label{fig:cs-mhd3}
\end{center}
\end{figure}

The key result of the 3D simulations is illustrated in Figures~\ref{3Dgammat40}-\ref{fig:cs-mhd3}.   Figure~\ref{3Dgammat40} shows the distribution of the Lorentz factor  in all three magnetic models by the end of the simulations, mostly in the longitudinal plane inclined to the x axis at $45^\circ$.   The solution for the unmagnetised jet is also shown as a reference model.  Figures \ref{fig:cs-mhd1}-\ref{fig:cs-mhd3}, complement these plots by showing the Lorentz factor distribution in the cross-sections $z=5,12,15$ and $40$.  The final time for all runs is $t=40$, which is one light crossing time of the domain in the $z$ direction. Since the flow velocity of the jet is almost 
equal to the speed of light, this is almost the same as the jet crossing time.

The structure of the jet in the model MHD1, where the strength of the magnetic field of the jet is the weakest, is almost same as the pure hydro model (HD). It exhibits transition from a laminar to a fully turbulent flow at around  $z=10$, which is somewhat upstream of the reconfinement point in the steady-state solution, which is located at $z \approx 23$ . The turbulence promotes entrainment of the external gas, mixing, and jet deceleration. The Lorentz factor reduces from $\Gamma=5$ down to $2\lesssim\Gamma \lesssim 3$.  

The recollimation instability also develops in the MHD2 model, but in contrast to MHD1 the transition to a turbulent state is not observed. A closer inspection of the flow structure in the jet cross-section (see figure \ref{fig:cs-mhd2}) reveals that the azimuthal number of the dominant mode gradually reduces from $m\approx20$ at $z=10$ to $m=4$ at $z=40$. Presumably, once the growth of higher order modes saturates, they get erased by numerical diffusion and the resultant flow with a thicker transition layer between the jet and the external gas can support only the modes of lowest order.  At   $z=40$, the non-linear $m=4$ mode clearly dominates other modes (figure \ref{fig:cs-mhd2}). One can see that it is aligned with the Cartesian grid, and this implies a strong bias due to the anisotropy of the numerical scheme.

In the MHD3 model, the magnetic field is sufficiently strong to completely suppress the recollimation instability. The shape of the jet cross-section shows a gradual transformation from the initial circular geometry to a square-like one near the far end of the computational domain. This square is also aligned with the Cartesian grid, which again suggests the numerical nature of this deformation. This numerical effect is likely to be exacerbated by the fact that the jet radius strongly reduces at the reconfinement point.

\section{Discussion}
\label{DISCUSSION}

\subsection{The nature of the instability}

Several researches interpreted the recollimation instability as a particular form of the Rayleigh-Taylor instability \citep{MatsumotoMasada2013,Matsumoto17,TKP17,Gottlieb-20a}.  In order to understand the arguments in favour of this interpretation,  it is perhaps most revealing to consider the evolution of the radial structure of a steady-state jet in an inertial frame moving with the jet speed along its axis. In this frame, the interface moves up and down relative to the jet axis and appears similar to the accelerated interface between two fluids in the problem considered by  \citet{Taylor:1950}.  Moreover, the structures formed at the nonlinear phase of the instability, but before the turbulent regime, are reminiscent of the fingers and bubbles characteristic to RTI. That is, if they are viewed in the jet cross section (e.g. see figures \ref{fig:cs-mhd1} and  \ref{fig:cs-mhd2}). 

\citet{MatsumotoMasada2013} assumed that the spatial oscillations of steady-state axisymmetric jets are equivalent to the temporal oscillations of under-expanded cylindrical ($\Pd{z}=0$) jets. They studied the temporal oscillations of such jets and observed an instability when the jet was heavier than the external gas \citep[cf. ][]{TKP17}. Hence they identified this instability as RTI. This identification makes sense, as in this problem both fluids are accelerated in the direction normal to the interface.  However, the similarity between the temporal oscillations of cylindrical jets and the spatial oscillations of steady-state jets is not sufficiently close to ensure the same nature of the instabilities observed in these problems.  In particular,  \citet{Gourgouliatos18a} have shown that the oscillating solutions for stationary jets are unstable not only when the jets are heavier than the external medium but also when they are lighter, and that in both these cases the instability looks the same. 

From the theoretical viewpoint, the key feature of the Taylor's setup is the same acceleration of both fluids in the direction normal to the interface between them. Hence in the non-inertial frame of the interface, the problem is identical to that studied by \citet{Rayleigh:1883}, where the initial steady-state configuration describes a hydrostatic equilibrium in gravitational field.
However, in the case of an oscillating steady-state jet this condition is not satisfied. Indeed, whereas the jet fluid moves along curved streamlines and hence experiences the centripetal  accelleration,  the external medium is at rest and hence has vanishing acceleration. Thus the jet problem is not a variant of the Taylor's problem. 

Locally, the motion of jet fluid along the curved interface between a steady-state oscillating jet and external gas is reminiscent of rotation.  \citet{Rayleigh:1917} established that rotating fluids may be subject to what is now known as the centrifugal instability (CFI).  Like RTI, this instability is local and hence the steady-state flow does not have to be a proper rotation \citep{Bayly:88}. In the plane normal to the streamlines, the structures produced by CFI may look very similar to fingers and bubbles of RTI \citep[e.g.][]{GK18}. However, in 3D they look more like ridges and trenches aligned with the flow streamlines.  This is exactly what was observed in the simulations of reconfined jets by \citet{Gourgouliatos18a} and is seen in our simulations as well. This suggests that the recollimation instability is an inertial instability closely related to the centrifugal instability of rotating fluids.

Using heuristic approach, \citet{GK18} derived a generalised Rayleigh instability criterion for relativistic rotating fluids. In the case of a discontinuity between two rotating fluids, it reduces to 
\begin{equation}
     [\Psi] < 0 \,,
\label{eq:Psi}
\end{equation} 
where $\Psi=w \Gamma^2 \Omega^2$, $\Omega$ is the angular velocity of rotation, and $[\Psi]=\Psi_o-\Psi_i$ where suffices ``i'' and ``o'' stand for the inner and outer sides of the  discontinuity respectively.   

In the Newtonian limit, $\Psi=\rho\Omega^2$ and the criterion reads  
\begin{equation}
     [\rho]\Omega_o^2 + \rho_i(\Omega_i+\Omega_o) [\Omega] < 0 \,. 
\label{eq:gCFI-critetion}
\end{equation} 
In the case of uniform density, this reduces to 
\begin{equation}
     [\Omega^2] <0 \,, 
\end{equation} 
which is the same  as the Rayleigh criterion for CFI in incompressible fluid. In the case of solid body rotation law ($[\Omega]=0$), (\ref{eq:gCFI-critetion}) reduces to 
\begin{equation}
     [\rho]<0 \,, 
\end{equation} 
which is the same as the instability criterion for RTI. Indeed,  in the frame rotating with the fluid, this case is equivalent to the equilibrium in the radial ``gravitational field'' with the acceleration $\bmath{g}=\Omega^2 \bmath{r}$, where $\bmath{r}$ is the radius vector of cylindrical coordinates aligned with the axis of rotation \citep[cf. ][]{ScHi18}.   

In the problem of a steady-state reconfined  jet, the external gas is at rest.  This corresponds to the rotation problem with $\Omega_o=0$. Hence, the instability criterion (\ref{eq:Psi}) reduces to
$$
-w_i\Gamma_i^2\Omega_i^2<0 \,,
$$ 
which is satisfied independently of the inertia of the external gas. This is in total agreement which is in agreement with results of jet simulations by \citet{Gourgouliatos18a}.  Thus we conclude that the recollimation instability is a variant of CFI.

\citet{KGM19} studied the role of axial magnetic field on the development of CFI at the interface between rotating relativistic fluids. Using heuristic approach, they concluded that the magnetic field suppresses the CFI modes with the wavelength below 
\begin{equation}
    \lambda_c = \frac{b^2}{w_1u_1^2 - w_2 u_2^2} r\sub{in} \,,
    \label{eq:lambda_c}
\end{equation}
where $b$ is the magnetic field strength as measured in the fluid frame, $w=\rho c^2 +(\gamma/\gamma+1)P$ is the relativistic enthalpy and $u=v\Gamma$, where $v$ is the flow rotational velocity and $\Gamma$ is the corresponding  Lorentz factor.  Indices ``1'' and ``2'' denote the fluids inside and outside of the interface with the curvature radius $r\sub{in}$ respectively. This criterion was in good agreement both with their Newtonian and relativistic  simulations.   Based on these results, they predicted complete suppression of CFI modes with the azimuthal wave number $m\geq 4$ at the interface between reconfined jets and external gas provided 
\begin{equation}
     \frac{\sigma}{1+\sigma} > \frac{(\theta_0 \Gamma)^2}{16} \,,
     \label{eq:sigma}
\end{equation}
where $\theta_0$ is the initial half-opening angle of the jet and $\Gamma$ is its Lorentz factor.

\begin{figure}
\begin{center}
\scalebox{0.9}{{\includegraphics{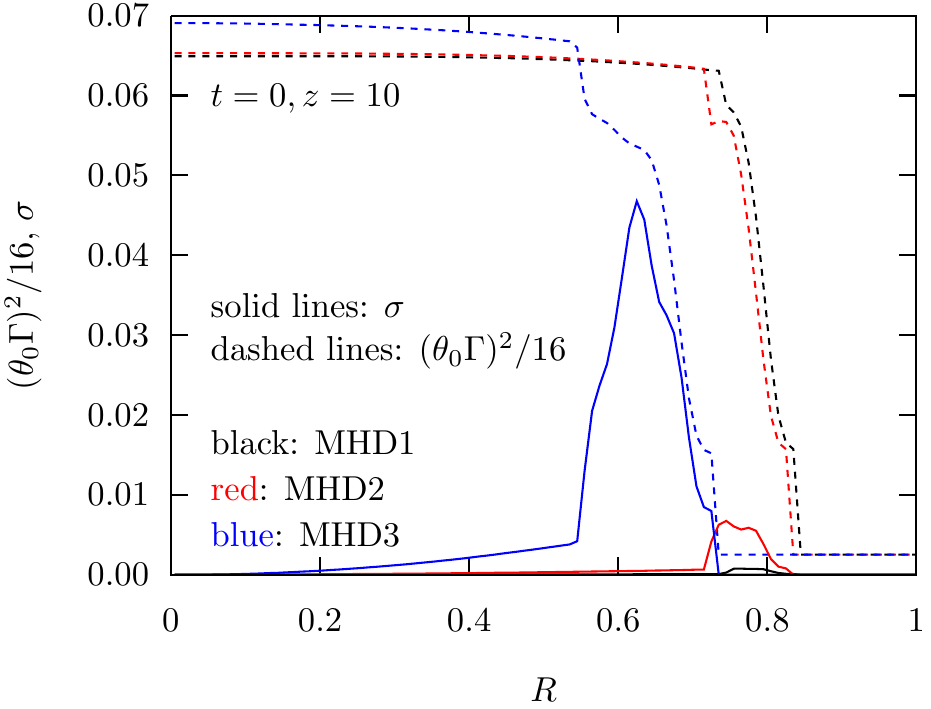}}} 
\caption{The radial distribution of ${(\theta_0\Gamma)^2}/16$ and $\sigma$ in the steady-state solutions.}
\label{sigma_criterion}
\end{center}
\end{figure}

Given the jet half-opening angle $\theta_0=0.2$ and the Lorentz factor $\Gamma=5$ of our jet models, equation \eqref{eq:sigma} predicts suppression of the recollimation instability for $\sigma>0.06$.  Given the maximum magnetisation of the jet plasma $\sigma\sub{max} \leq 0.01$ at the nozzle, one would expect the instability to develop in all three models. However, we find that in MHD3 with $\sigma\sub{max} = 0.01$ it is completely suppressed. 
One factor promoting this suppression is the increase of $\sigma$ at the reconfinement shock. Figure \ref{fig:sigma} shows that in the shocked layer the magnetisation increased up to $\sigma\approx 0.04$. Another factor is the decrease of the jet Lorentz factor at the interface with the external gas.  It is introduced already at the nozzle via boundary conditions, and downstream it is further amplified by the reconfinement shock.  Figure \ref{sigma_criterion} shows the radial distribution of $(\theta_0 \Gamma)^2/16$ and $\sigma$ at z=10 in all magnetic models. One can see that the suppression criterion is not satisfied for the shocked layer in the MHD1 and MHD2 models, but it is marginally satisfied in the MHD3 model.  Thus the results of our jet simulations are in a good agreement with the predictions based on the study of magnetic CFI in rotating fluids.

When we were working on this paper,  the results of a related numerical study of relativistic jets was published by  \citet{Gottlieb-20b}.  They also investigated the impact of magnetic field on the recollimation instability of relativistic jets, but in the context of gamma ray bursts (GRB), and concluded that $\sigma \geq 10^{-2}$ leads to a suppression of the instability.   Apparently, they were unaware of our study of the magnetic CFI in rotating flows and did not compared their results with the criterion \eqref{eq:sigma}.  

When comparing with the criterion \eqref{eq:sigma}, one have to keep in mind that they consider relativistically hot jets, which can be thermally accelerated.  Indeed, the thermal acceleration of weakly magnetised jets leads to $\Gamma \propto R_j$ where $R_j$ is the jet radius \citep[e.g.][]{Kom-11}. Hence, the flow Lorentz factor may significantly increase compared to its value at the nozzle.  Although  \citet{Gottlieb-20b} do not show the variation of jet Lorentz factor, this cannot be large as the jet radius prior to the reconfinement point  increases only slightly in models for long GRB and only by a factor of few in their setup for short GRBs.  At the nozzle, they set $\Gamma_0 \theta_0 = 0.7$, which is only slightly below of   $\Gamma \theta_0 = 1$ in our simulations. Thus, at least in the case of long GRBs, their results are consistent with ours and with the criterion \eqref{eq:sigma}.

\subsection{Implications to AGN jets}

According to the VLBI observations of AGN jets, the mean value of $\theta_j \Gamma$ is about 0.2 \citep{Jorstad-05,PKLS-09,C-B-13}.  For such jets, the criterion \eqref{eq:sigma} gives the critical value $\sigma\sub{cr} \approx 0.0025$.  Such a small value is very problematic for the magnetic collimation-acceleration mechanism because it dramatically loses efficiency when $\sigma$ drops to the value about unity.  Based on the asymptotic solutions, \citet{lyub-10} gives $\sigma=1$ at the distance $z\sub{1} = 10^2-10^3 r_g$ (see also \citet{KBVK-07}) and $\sigma=0.1$ at the distance $z\sub{0.1} = z\sub{1} \Gamma_{max}^4$ from the central black hole, where $r_g$ is its gravitational radius and $\Gamma_{max}$ is the terminal jet Lorentz factor. Repeating his calculations for $\sigma=0.01$, we find $z\sub{0.01} = z\sub{1} \Gamma_{max}^{49}$ (the power index of 49 is not a typo), which is ridiculously high for any realistic value of $\Gamma_{max}$.  Hence, if AGN  jets are accelerated via this mechanism then their magnetisation never becomes small enough to allow the recollimation instability. 

Given the jet power and radius, one can relate the strength of its magnetic field with $\sigma$.  Assuming a uniform jet, its total power can be found as 
$$
L = \left(w+\frac{B\sub{cm}^2}{4\pi}\right)\Gamma^2 c \pi R_j^2 
$$           
and hence 

\begin{equation}
B\sub{cm}^2 = \frac{4L}{c\Gamma^2 R_j^2} \frac{\sigma}{\sigma+1} \,.
\label{eq:b-sigma}
\end{equation}
We apply this result to the M87 jet, the best studied case of all AGN jets.  It has been suggested that its optical knot HST1 coincides with the reconfinement point, the location where the reconfinement shock reaches the jet axis \citep{SAK-06,Nalewajko-12}. It is located at about 250 pc from the supermassive black hole of M87 \citep{BSM-99}. At around this point, the radio observations suggest a transition from acceleration to deceleration of the jet \citep{Asada-14}, as well as a transition from parabolic to conical geometry \citep{AN-12}.           
The jet radius at the deprojected distance $z=100\,$pc  is $R_j\approx 2\,$pc \citep{AN-12} and the Lorentz factor $\Gamma \approx 5$ \citep{Asada-14}. The mean power of the M87 jet can be estimated based on the work done by their expanding radio lobes against the thermal pressure of the surrounding them X-ray emitting gas. This yields $L\approx 10^{44}\,$erg/s \citep{OEK-00}. Substituting these values into  \eqref{eq:b-sigma} we find 
$$
B\sub{cm}\approx 0.11 \fracp{\sigma}{0.001}^{1/2}\mbox{mG} \,.
$$ 
This is well below the equipartition value for the HST1 knot $B\sub{eq}\approx 1\,$mG \citep{HBJ-03} and the value based on the variability time-scale of its emission during flares, $B\sub{var}\approx 0.6\,$mG \citep{HCS-09}.  In fact, these  observational estimates suggest  $\sigma = 0.03\,-\,0.08$, which is well above  $\sigma\sub{cr}=0.0025$\footnote{Note that the data for the M87 jet suggest $\theta_j \approx 0.02$ and $\Gamma \theta_j \approx 0.1$ at $z=100\,$pc,  in agreement with the data for other VLBI jets.}. 

If the magnetisation of AGN jets drops well below $\sigma=0.1$ at sub-kpc scales, then their physics is significantly more complicated than it is assumed in the collimation-acceleration  model.  In fact, there are several indications that this may be the case. For example, the rich morphology of AGN jets is very different from the featureless structure of the theoretical model. The superluminal motion in  pc-scale jets and the flares of their cores are clear manifestations of the central engine variability.  The synchrotron optical and X-ray emission of AGN jets require in-situ particle acceleration, which suggests a dissipation of either the kinetic energy of the jet bulk motion or of its magnetic energy \citep[see][and references therein]{MBB-20}. The polarisation of the jet emission indicates the presence of a strong longitudinal component, which is not expected in the ideal model as $B_z\propto R_j^{-2}$ and $B_\phi \propto R_j^{-1}\Gamma^{-1}$. In particular, the longitudinal component is prevalent upstream of the HST1 knot of M87 jet \citep{PBZ-99}.       

There exists an alternative model of the jet acceleration, where it is powered by the energy released via magnetic dissipation \citep[e.g.][]{SDD-01}. In this model, the dissipated magnetic energy is converted into heat, and as the jet expands the heat is converted into the kinetic energy of the jet.   Such magnetic dissipation may occur even in freely expanding (unconfined) jet, provided its magnetic field frequently changes polarity due to some magnetic activity of the jet engine. Suppose that the engine changes polarity on the time scale $\Delta t_e$. Then the jet contains blocks of alternating azimuthal magnetic field of the length $l_e=c \Delta t_e$ in the engine (observer) frame. In the jet frame their length is $l'_e = \Gamma l_e=\Gamma c \Delta t_e$. If $v\sub{in}$ is the reconnection speed, then the time scale of magnetic dissipation $\Delta t'_d \approx l'_e/v\sub{in} \approx \Gamma \Delta t_e/\beta\sub{in}$.  In the observer frame, the corresponding time is $\Delta t_d \approx \Gamma^2 \Delta t_e$, leading to the characteristic length scale of magnetic dissipation 
\beq
l_d \approx \Gamma^2 c \Delta t_e/\beta\sub{in} \,.
\eeq   
Based on  numerical simulations (PIC) of relativistic pair  plasma \citet{Liu-15} give 
\beq
    v\sub{in} \approx 0.1 c_a \fracp{1+\sigma}{ 1+0.01\sigma}^{1/2}  
\eeq
where 
$$
c_a=c\fracp{\sigma}{1+\sigma}^{1/2}
$$
is the Alfv\'en speed. Using $\sigma=\Gamma=5$, and $\Delta t_e=1\,$yr (assuming that the polarity changes on the same time scale as the ejection of new superluminal components in VLBI jets), we obtain $l_d \approx 30\,$pc. This shows that, the large-scale azimuthal magnetic field can be destroyed well before the typical reconfinement scale of FR-1 jets.  
We envisage that at $z \gg l_d$, the jet contains mostly small-scale (tangled or turbulent) magnetic field  and the plasma magnetisation drops down to $\sigma <1$.   Even if the magnetisation is not as low as $\sigma < \sigma\sub{cr}$ the recollimation instability may still develop if this $\sigma$ is attributed almost entirely to the small-scale field.

If however this scenario is not followed by the AGN jets and the recollimation instability is not the reason for the observed flaring and deceleration of FR-1 jets within the framework of this paradigm, then KHI and CFI may be the ``culprits'' instead. In this regard, it is intriguing that MHD3 model shows no signs of these instabilities. Presumably the magnetic field is sufficiently strong to suppress KHI and not strong enough to promote a sufficiently rapid growth of CDI. The shear layer is also known to inhibit these instabilities in cylindrical jets \citep[e.g.][]{MPG-16,Kim-16}.  The non-cylindrical structure of our jets may play a role too.

Future observations with ngVLA  are expected to allow detailed study of some AGN jets on the reconfinement scale and observationally explore their stability properties on this scale \citep{LKK-18,PBM-19}.   Numerical simulations can be used to explore the reconfinement dynamics of jets with $\sigma\approx 0.1$ and that of multi-component jets.  

If the AGN jets are almost certainly produced by magnetic central engines, the jets of gamma-ray bursts (GRB)  may well be neutrino-driven and as the result have much lower magnetisation than the AGN jets \citep{Woosley-93,MacF-99}.  Hence the recollimation instability is likely to be important for these jets.

\section{Conclusion}
\label{CONCLUSION}

Recollimation of astrophysical jets can lead to instability. This recollimation instability is powered by the centrifugal force emerging along the curved streamlines of recollimating jets and is a variant of the classic centrifugal instability of rotating fluids.  Many types of astrophysical jets are magnetised and strong regular magnetic fields can be the most important component of their ``jet engines''.  In this study, we explored the role played by such regular magnetic in the development of the recollimation instability. 

As an example, we considered the reconfinement of initially free-expanding jets with purely azimuthal (toroidal) magnetic field by the thermal pressure of external gas, using the parameters suitable to the so-called ``naked'' AGN jets.  In the case of unmagnetised jets, we find that the recollimation instability leads to fully-developed turbulence soon after the reconfinement point, entrainment of the external gas, and rapid deceleration of the jets. This is in agreement with the previous studies of such jets, which have lead to the suggestion that the instability may be responsible for the  observed morphology of FR-1 extragalactic radio sources.  

However, we find that even a rather weak azimuthal magnetic field can fully suppress the development of this instability. For the jets with the half-opening angle $\theta_0=0.2$ and the Lorentz factor $\Gamma=5$, the critical relativistic magnetisation parameter can be as low as $\sigma\sub{cr}=0.01$.            

These results are in good agreement with the predictions based of the results for magnetic centrifugal instability of rotating flows, which relate $\sigma\sub{cr}$ to the product $\theta_0\Gamma$. On one hand, this confirms the identification of the recollimation instability as a variant of the (local in nature) centrifugal instability. On the other hand, this allows us to extrapolate the results to the regimes typical to parsec-scale AGN jets where the observations suggest $\theta_0\Gamma\approx 0.2$, and estimate their critical magnetisation as $\sigma\sub{cr} \approx 0.002$.    Such a low magnetisation can not be reached on the scales typical for AGN jets if they are accelerated via the ideal magnetohydrodynamic collimation-acceleration mechanism. In this is indeed the case, the observed disruption of FR-1 jets must have a different origin.

If however, the regular azimuthal magnetic field of AGN jets is destroyed before the jet disruption, then the recollimation instability may still be relevant. For example, the jet engine may change its magnetic polarity on a regular basis, leading to a striped magnetic structure of the jets.  This creates conditions for magnetic reconnection at the interfaces between stripes with opposite direction of magnetic field. Provided the characteristic time scale of this variability of the central engine is the same as for the ejection of superluminal component ($\approx$ one year), the reconnection may indeed be completed before kpc scales, leaving behind mostly small-scale field.  In fact, this may explain why the polarisation observations of AGN jets are often inconsistent with the predominantly azimuthal magnetic field.  The magnetic dissipation accompanying the reconnection may power the particle acceleration required to explain the observed emission of the jets.


\section*{Acknowledgments}
Jin Matsumoto and Serguei Komissarov were supported by the STFC grant No. ST/N000676/1. Part of the numerical simulations were carried out on the STFC-funded DiRAC I UKMHD Science Consortia machine, hosted as part of and enabled through the ARC HPC resources and support team at the University of Leeds (www.dirac.ac.uk). Another part of the numerical simulations were carried out on Cray XC50 at the Center for Computational Astrophysics and National Astronomical Observatory of Japan and Cray XC40 at YITP in Kyoto University. Jin Matsumoto was also supported by Research Institute of Stellar Explosive Phenomena at Fukuoka University and the associated project (No. 207002), and
also by JSPS KAKENHI Grant Number (JP19K23443 and JP20K14473).

\section*{Data Availability}

The data underlying this article will be shared on reasonable request to the corresponding author.

\bibliographystyle{mnras}
\bibliography{papers,jets,num,astro}


\end{document}